\documentclass[sigconf]{acmart}
\AtBeginDocument{%
  }

\setcopyright{acmlicensed}
\copyrightyear{2018}
\acmYear{2018}
\acmDOI{XXXXXXX.XXXXXXX}
\acmConference[Conference acronym 'XX]{Make sure to enter the correct
  conference title from your rights confirmation email}{June 03--05,
  2018}{Woodstock, NY}
\acmISBN{978-1-4503-XXXX-X/2018/06}

\usepackage{booktabs,tabularx}
\usepackage{graphicx}
\usepackage{subcaption}
\usepackage{multirow}

\usepackage{amsmath}
\usepackage{mathtools}

\title{Rotate Both Ways: Time-and-Order RoPE for Generative Recommendation}

\author{Xiaokai Wei}
\email{xwei@roblox.com}
\affiliation{\institution{Roblox} \city{San Mateo, California} \country{USA}}
\author{Jiajun Wu}
\email{jiajunwu@roblox.com}
\affiliation{\institution{Roblox} \city{San Mateo, California} \country{USA}}
\author{Daiyao Yi}
\email{dyi@roblox.com}
\affiliation{\institution{Roblox} \city{San Mateo, California} \country{USA}}
\author{Reza Shirkavand}\authornote{Contribution was made while being a graduate collaborator at Roblox.}
\email{rezashkv@umd.edu}
\affiliation{\institution{University of Maryland} \city{College Park, Maryland} \country{USA}}
\author{Michelle Gong}
\email{mgong@roblox.com}
\affiliation{\institution{Roblox} \city{San Mateo, California} \country{USA}}

\renewcommand\shortauthors{Wei et al.} % short author line

\begin{abstract}
Generative recommenders, typically as transformer-based autoregressive models, predict the next item/action from a user’s interaction history. Their effectiveness relies on how the model represents \emph{where} an interaction event occurs in the sequence (discrete index) and \emph{when} it occurred in wall-clock time. Prevailing approaches inject time via learned embeddings or relative attention biases. In this paper, we argue RoPE-based approaches, if designed properly, could be a stronger alternative for jointly modeling the temporal and sequential information in user behavior sequences. 

While vanilla RoPE in LLM only considers the index/order of tokens, generative recommendation requires to incorporate both event time and token index. To tackle the challenges, we propose \emph{Time-and-Order RoPE} (TO-RoPE), a family of rotary position embedding (RoPE) designs that treat index and time as angle sources that shape the geometry of $Q,K$ directly. In particular, three instantiations of this framework are presented and discussed: early fusion, split-by-dim and split-by-head. We perform extensive experiments on both publicly available dataset and proprietary industrial dataset, and show TO-RoPE variants consistently improve Hit@K/NDCG@K over existing approaches to encoding time and index. Our results position rotary embedding as a simple, principled, and deployment-friendly foundation for generative recommendation.
\end{abstract}


\received{20 February 2007}
\received[revised]{12 March 2009}
\received[accepted]{5 June 2009}

\begin{document}
\maketitle

%===================================================
\section{Introduction}

Modern recommender systems increasingly rely on user behavior sequences for enabling higher-quality recommendations and more precise, timely personalization \cite{Zivic2024SR, Zhang2025KillingTB, Zhou2025OneRecTR, Kang2018SasRec, Zhai2024ActionsSL, Chen2025PinFM, Xia2023TransActTR}. Modeling behavior as a sequence is appealing because it captures evolving intent, contextual dependencies, and cross-item transitions that static user/item features cannot. For example, in e-commerce, the ordered pattern \emph{view} $\rightarrow$ \emph{compare} $\rightarrow$ \emph{add-to-cart} is far more predictive of a purchase than any single action in isolation. In streaming, the trajectory \emph{S1E1} $\rightarrow$ \emph{S1E2} $\rightarrow$ \emph{S1E3} strongly implies \emph{S1E4} next, whereas unrelated popular titles might be recommended without utilizing the sequential pattern.

 \begin{figure}[t]
 \centering
  \includegraphics[width=1\columnwidth]{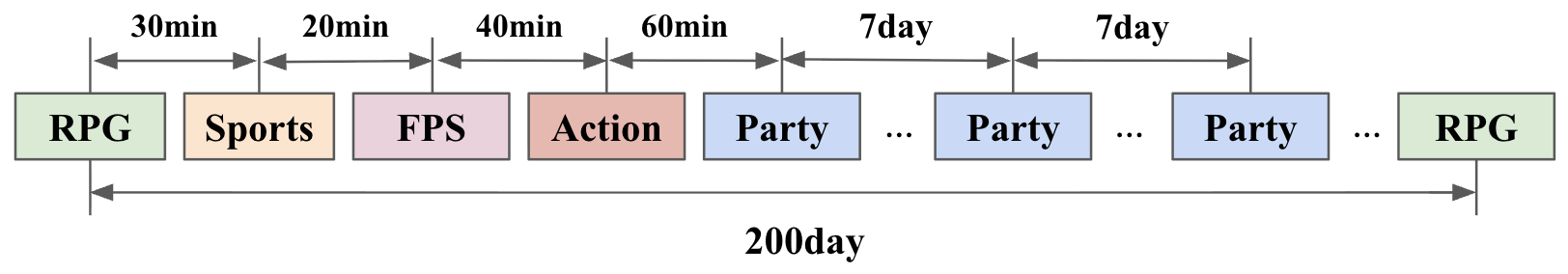}
 \caption{Illustration of typical user behavior sequence. Several time patterns conoccur, including short-term burstness, long-term patterns and periodicity.}
 \label{fig:user_behavior}
 \end{figure}

As Transformer-based architectures demonstrate strong scaling behavior across parameters and data in diverse domains (LLMs, vision) \cite{Raffel2019ExploringTL, Peebles2022ScalableDM}, recent sequence modeling methods has largely converged on Transformers, especially in generative recommendation (GR). Analogous to language modeling, GR (e.g., HSTU \cite{Zhai2024ActionsSL} and related models \cite{Han2025MTGR}\cite{Zhou2025OneRecTR}\cite{Wei2025actionitemcoupling}) typically adopts a next-item prediction objective, treating user histories as token sequences and items as the vocabulary. This formulation scales naturally with data, benefits from well-optimized attention kernels, and unifies training and inference in an autoregressive manner.

Despite the analogy to LLM training, a key distinction between GR and LLM is that user behavior sequences exhibit structure along \emph{two} axes: discrete order (index) and occurrence time. Unlike text, user interactions arrive irregularly with gaps, session boundaries, and calendar effect. In practice, three temporal/sequential patterns usually arise:
\begin{itemize}
\item \textbf{Short-term burstiness.} Users generate dense micro-sessions (e.g., binge multiple shows/videos or try multiple games within hours) separated by long inactive periods.
\item \textbf{Long-term rhythms.} Interests resurface after weeks or months (e.g., returning to a franchise or genre previously enjoyed), requiring sensitivity to distant yet relevant history.
\item \textbf{Temporal periodicity (calendar effects)} User behavior often aligns with time-of-day/day-of-week/season (e.g., party games on Saturday nights, commute playlists on weekday mornings), demanding phase-aware modeling beyond monotone recency.
\end{itemize}
\noindent An illustrative user history sequence in gaming domain is shown in Figure~\ref{fig:user_behavior}. Capturing these patterns reliably requires positional encodings that incorporate order and time jointly and effectively. Existing sequence modeling techniques often inject time via heuristics such as time gaps \cite{Li2020TimeIA, Chen2022TimeLA}, simple temporal embeddings \cite{Zhou2025OneRecTR} concatenated to item tokens, or relative time biases \cite{Si2024TWINV2} \cite{Zhai2024ActionsSL} added to attention scores. These choices are easy to implement but fragile: they often require manual feature engineering, over-impose inductive bias and struggle to capture calendar effects.

An ideal encoding should readily handle these temporal/sequential characterisitics of user behavior without heavy reliance on handcrafted time features (e.g., fixed buckets for time gaps), instead letting the model learn appropriate timescales for effective information utilization. In this paper, we argue Rotary Position Embedding (RoPE) \cite{Su2021RoFormerET} based methodology provides a versatile and solid foundation for meeting these criteria. Nevertheless, directly applying vanilla RoPE to recommendation which treats user logs as a purely ordered sequence and ignores time, would be suboptimal for irregular, bursty, and periodic user behavior. The open question is therefore: how should we coordinate discrete order and time within rotary embeddings so they act in synergy rather than competition?

To address the challenges, we propose a unified framework and methodology: Time-and-Order RoPE (\textbf{TO-RoPE}) that incorporates both time and order/index as angle sources for rotary embeddings. Our general formulation supports multiple instantiations: (a) \textbf{early fusion} of time and index in the angle domain with learnable gates and timescales, and two practical specializations that improve stability and interpretability: (b) \textbf{split-by-head}, where some heads rotate by time and others by index (head-level gating), and (c) \textbf{split-by-dimension}, where some rotary planes within a head use time while others use index (dimension-level gating). These designs preserve standard attention kernels (Flash-Attention \cite{Dao2022FlashAttentionFA, Dao2023FlashAttention2FA} compatible) and add neglible additional parameters. Together, these factors position TO-RoPE not merely as a positional add-on, but as a principled geometric prior for unifying time and order in generative recommendation. 

Our contributions can be summarized as follows:
\begin{itemize}
\item \textbf{Criteria.} We articulate the desiderata for positional encodings in generative recommendation (burstiness, long-range rhythms, periodicity), and discuss where existing heuristics fall short and why RoPE offers strong promise on these aspects.
\item \textbf{Effective Approaches with a Unified View} Reconizing the limitation of either index-only RoPE or time only RoPE, we design a \emph{Time-and-Order RoPE} family that integrates time and order with a unified view. Also, we discuss in details three instantiations (early fusion, split-by-head and split-by-dim) and their characteristics. 
\item \textbf{Evaluation.} We conduct extensive experiments on both public (e.g., MovieLens-20M) dataset and proprietary large-scale industrial dataset. In addition, we study the tradeoff between different capacity allocation strategies to provide practical guidance. We also include attention analysis to help explain why RoPE-based designs better capture user behavior.
\end{itemize}

\section{Related Work}

\subsection{Incorporate Time in User History Modeling}

In recommender systems, temporal patterns usually plays an important role in modeling user behavior sequence. Existing sequential recommenadtion models often explicitly construct time interval features to capture the temporal information \cite{Li2020TimeIA, Chen2022TimeLA, Shirkavand2025CatalogNativeLS, Zhang2022ATS,  Chen2021ExploringPA, Si2024TWINV2}. For example, LONGER \cite{Chai2025LONGER} and TransAct \cite{Xia2023TransActTR} directly encode timestamp as embedding which is used as input to the transformer models. \cite{Si2024TWINV2} and \cite{Li2020TimeIA} incorporate time interval into attention bias, similar to the idea of T5-bias \cite{Raffel2019ExploringTL} or ALiBi \cite{Press2021TrainST}. Recently, a growing line of work in recommendation focuses on a generative paradigm. Recent progress in this direction \cite{Zhai2024ActionsSL, Han2025MTGR, Rajput2023RecommenderSW, Chen2025PinFM, Zhou2025OneRecTR, Shirkavand2025CatalogNativeLS} demonstrates strong potential of this paradigm: scaling up in parameter sizes significantly improves the model prediction accuracy. On positional encoding, HSTU relies on relative bias calculated from time and index; Onerec \cite{Zhou2025OneRecTR} uses learnable position embedding as model input.

\subsection{Position Embedding for LLMs}
Positional encoding is essential for transformer-based models as the multi-head self-attention \cite{Vaswani2017AttentionIA} is permutation-invariant. Early absolute schemes include fixed sinusoidal embeddings \cite{Vaswani2017AttentionIA} or learneable absolute embeddings \cite{Radford2019LanguageMA}, which might extrapolate poorly and overfit to the maximum context seen in training. Relative methods condition attention on token distance rather than index, and for instance, additive relative bias matrices capture pairwise offsets. Efficient variants include T5-bias \cite{Raffel2019ExploringTL} and ALiBi \cite{Press2021TrainST}, which adds head-dependent linear slopes to the logits and yields strong long-context extrapolation with negligible overhead. Table \ref{table:compare_pe} summarizes the charateristics of representative positional embedding methods.

Among alternatives, Rotary Position Embedding (RoPE) \cite{Su2021RoFormerET} rotates each even–odd channel pair of $Q,K$ by an angle linear in position, turning inner products into relative phase comparisons. RoPE is parameter-light, extrapolates well \cite{Peng2023YaRNEC}, and preserves compatibility with optimized attention kernels. Beyond text, RoPE has been adapted to vision \cite{Heo2024RotaryPE} by applying rotations along spatial axes and to video \cite{Gao2024TCLLaVART} \cite{Wei2025VideoRoPEWM} by extending rotations over spatial–temporal dimensions.

% Requires: \usepackage{booktabs,tabularx}
\begin{table*}[t]
\centering
\small
\begin{tabularx}{\linewidth}{@{}l l l X X@{}}
\toprule
\textbf{Method} & \textbf{Type} & \textbf{Key Feature} & \textbf{Pros} & \textbf{Cons} \\
\midrule
Sinusoidal \cite{Vaswani2017AttentionIA} & Absolute (fixed) & Fixed sin/cos of position &
No learned params; extrapolates to long sequences &
Not data-adaptive; limited expressivity \\
Absolute Learnable \cite{Radford2019LanguageMA} & Absolute (table) & Learned embedding per index &
Flexible; fits task distributions &
Poor extrapolation beyond max length; larger memory \\
T5 Bias \cite{Raffel2019ExploringTL} & Relative (bucketed) & Learned attention bias by distance &
Captures relative distances; lightweight &
Heuristic buckets; affects logits only \\
ALiBi \cite{Press2021TrainST} & Relative (linear) & Head-specific linear bias to scores &
No table; strong long-context extrapolation &
Low expressivity; no periodic patterns \\
RoPE \cite{Su2021RoFormerET} & Relative (rotary) & Rotate \(Q/K\) planes by angle (phase) &
Extrapolates well; multi-scale; enables time+order fusion &
Needs scale calibration; less interpretable than biases \\
\bottomrule
\end{tabularx}
\caption{Representative positional encoding methods.}
\label{table:compare_pe}
\end{table*}

%===================================================
\section{Preliminaries}
\subsection{Rotary Position Embeddings (RoPE)}
Given input sequence embeddings $X \in \mathbb{R}^{T \times d_{\text{model}}}$ (with $T$ time steps and model width $d_{\text{model}}$), Multi-Head Attention (MHA) \cite{Vaswani2017AttentionIA} computes per-head projections  (h=1,\dots,H):
\[
Q_h = X W^Q_h,\quad K_h = X W^K_h,\quad V_h = X W^V_h. 
\]
with $W^Q_h,W^K_h,W^V_h\in\mathbb{R}^{d_{\text{model}}\times d}$ and head width $d$ (even). Each head computes
\[
\mathrm{Attn}_h(X)=\mathrm{softmax}\!\Big(\tfrac{Q_hK_h^\top}{\sqrt{d}}\Big)\,V_h,
\qquad
\mathrm{MHA}(X)=\Big[\;\Vert_{h=1}^{H}\mathrm{Attn}_h(X)\Big]W^O.
\]

RoPE injects position by \emph{rotating} the even–odd channel pairs of $Q_h$ and $K_h$ in each head, while $V_h$ is unchanged \cite{Su2021RoFormerET}. For plane $k\in\{0,\ldots,\tfrac{d}{2}-1\}$ and position $i$,
\begin{equation}
\label{eq:rope-rotate}
\begin{bmatrix}q'_{i,2k}\\ q'_{i,2k+1}\end{bmatrix}
=
\underbrace{\begin{bmatrix}\cos\theta_{i,k} & -\sin\theta_{i,k}\\ \sin\theta_{i,k} & \cos\theta_{i,k}\end{bmatrix}}_{R(\theta_{i,k})}
\begin{bmatrix}q_{i,2k}\\ q_{i,2k+1}\end{bmatrix},
\quad \\
\begin{bmatrix}k'_{j,2k}\\ k'_{j,2k+1}\end{bmatrix}=R(\theta_{j,k})
\begin{bmatrix}k_{j,2k}\\ k_{j,2k+1}\end{bmatrix}.
\end{equation}
With vanilla RoPE (index-only), the rotation angle is linear in the discrete position $i$:
\[
\theta^{p}_{i,k}=i\,\omega^{p}_k,
\]
so each plane contributes a relative-phase term to the score,
\begin{equation}
\label{eq:rope-relphase}
\langle q'_i,k'_j\rangle \;\propto\; \sum_{k=0}^{\frac{d}{2}-1}\cos\!\big((i-j)\,\omega^{p}_k\big),
\end{equation}
so attention depends on \emph{relative} index gaps $\Delta i=i-j$. Because $\{\omega^{\mathrm{pos}}_k\}$ spans low$\to$high frequencies, RoPE simultaneously supports long-range ($\Delta i$ large; low $\omega$) and local ($\Delta i$ small; high $\omega$) relations.

RoPE uses a fixed, log-spaced bank of per-plane frequencies.
\begin{equation}
\label{eq:freq-bank}
\omega^{p}_k \;=\; \text{base}^{-\frac{2k}{d}},\qquad k=0,\ldots,\tfrac{d}{2}-1.
\end{equation}
where  $\text{base}$ is typically $10000$. The same ladder is shared by $Q_h$ and $K_h$ within a head and typically across heads, ensuring consistent phase geometry before the softmax.

\subsection{Rotary Embeddings: From Index to Time}
\label{subsec:time-only-rope}
Vanilla RoPE encodes discrete index $i$, but for modeling user sequences in recommendation we often need a geometry over continous time. In this subsection, we discuss how to adapt and base RoPE calculation on the time sequence of interaction events. This requires to replace the index angle with a time angle so attention depends on \emph{relative time}.

Let raw timestamps be datestrings or Unixtime $u_i$. Normalize the raw timestamp $u_i$ to a scalar (e.g., seconds or days depending on particular application):
\[
\tau_i=\frac{u_i-u_{\mathrm{ref}}}{s}
\]
where $u_{\mathrm{ref}}$ is a chosen anchor (e.g., global date) and $s$ is normalization factor to make the time scalar in reasonable range (e.g., similar order of magnitude as index range). This produces a one-to-one aligned sequence $[\tau_i]$ for the user’s interactions. In rotary embeddings, since attention depends on relative time $\Delta\tau = \tau_i-\tau_j$, it is timezone-invariant and anchor-invariant.

Define the time angle
\[
\theta^t_{i,k}=\tau_i\,\omega^t_k \quad \text{or} \quad \theta^t_{i,k}=\tau_i\,\alpha^t_k\,\omega^t_k,\quad \alpha^t_k>0.
\]
The effective period per plane is then $\lambda^{\mathrm{t}}_k=\frac{2\pi}{\omega^t_k}$. This allows direct inspection of learned  bands that correponds to certain periodicity (e.g., day/week/seasonal). For a head with $N=\tfrac{d}{2}$ planes and time frequency bank $\{\omega^t_k>0\}_{k=1}^{N}$,
\[
\theta^t_{i,k} \;=\; \tau_i\,\omega^t_k,\qquad
\Delta\theta^t_{ij,k} \;=\; (\tau_i-\tau_j)\,\omega^t_k \;=\; \Delta\tau_{ij}\,\omega^t_k,
\]
and the plane-$k$ contribution behaves like a time kernel
$\cos(\Delta\tau_{ij}\,\omega^t_k)$.
High $\omega^t_k$ emphasizes short gaps (bursts); low $\omega^t_k$ captures long-range recency. A head’s similarity can be written as follows:
\[
S(\Delta\tau)\;\propto\;\sum_{k} w_k\,\cos\!\big(\Delta\tau\,\omega^t_k\big),
\]
so two interactions \emph{harmonically align} when their gap $\Delta\tau$ places them at similar phases across relevant bands—e.g., near multiples of the daily period $2\pi/\omega_{\text{day}}$ and the weekly period $2\pi/\omega_{\text{week}}$. This yields large, constructive contributions and a high dot product in $Q/K$ space.

Similar to index-only RoPE, one need to keep $\omega_k$ monotone (log-spaced) to avoid band swaps; ensure $\omega_{\min}$ covers the longest rhythms and $\omega_{\max}$ responds to session bursts. Two ways can be considered to build $\{\omega^t_k\}$:
\begin{enumerate}
\item \textbf{Base-form:} 
$\omega^t_k=\text{base}_{t}^{-\frac{2(k-1)}{d}}$ with $\text{base}_t\!\in\![10^3,10^5]$; add $\alpha^t_k$ to adapt magnitudes to your units.
\item \textbf{Wavelength-targeted:} choose periods $\lambda\in[\lambda_{\min},\lambda_{\max}]$ in the units of $\tau$ (e.g., seconds or days), set
\[
\omega_{\max}=\frac{2\pi}{\lambda_{\min}},\quad \omega_{\min}=\frac{2\pi}{\lambda_{\max}},
\]
and log-space between them:
$\omega^t_k=\omega_{\max}\big(\tfrac{\omega_{\min}}{\omega_{\max}}\big)^{\frac{k-1}{N-1}}$.
\end{enumerate}

\subsection{Why RoPE Fits Generative Recommendation}
Now we discuss why RoPE could be a stronger alternative to existing methods in generative recommendation. RoPE rotates $(Q,K)$ by angles so similarity depends on relative phase across planes. This re-shapes feature geometry before softmax. Relative-bias methods (e.g., T5/ALiBi/HSTU-like) add terms to logits only, and absolute PEs add vectors that memorize positions without a clean relative notion. Besides, relative-bias typically needs separate tuned biases for time and index and cannot alter $Q/K$ geometry; absolute schemes require concatenating engineered time features (gaps, TOD/DOW) and relying on MLPs to learn interactions.

RoPE provides a frequency ladder (high$\rightarrow$low) and each rotary plane acts as a band-pass filter over time (and/or index), so with more data the model allocates capacity across frequencies to fit bursts, rhythms, and periodic cycles—rather than relying on hand-crafted, monotone biases (e.g., “longer gap $\Rightarrow$ larger penalty”). This matches scaling behavior: as datasets grow, gradient descent discovers learns which bands and phases matter from data, not from fixed decay curves. Relative-bias is mostly monotone in distance (linear/bucketed), weak for periodicity; absolute PEs depend on coarse binning of time.

Additionally, RoPE is intrinsically \emph{relative} and extrapolates to longer contexts \cite{Peng2023YaRNEC}, absolute learned tables degrade beyond the trained max index/time; relative-bias extrapolates better but with a fixed functional form (e.g., linear decay), limiting multi-scale expressivity.

\subsection{Limitations of Single Source RoPE}

Though RoPE provides a solid foundation for modeling temporal or sequential information, single source RoPE (i.e., index-only or time-only) is likely to be suboptimal for generative recommendation, as explained in the following examples.

\subsubsection*{Why Index-Only RoPE is insufficient}
Index-only RoPE enforces ordinal locality but ignores staleness and temporal periodicity. Incorporating wall-clock time is essential for robust and context-aware recommendation.

\paragraph{Staleness vs.\ adjacency (close in index but stale).}
\begin{itemize}
\item \textbf{E-commerce.} A user \emph{adds-to-cart} a jacket, disappears for 3 weeks, then returns and \emph{buys} electronics. Index-only ($\Delta i{=}1$) over-weights the jacket, despite a long gap indicating intent drift.
\item \textbf{Short-form video.} A user watches a cooking clip, closes the app, and after 48 hours opens a skateboarding clip. Index-only over-credits the cooking clip due to index adjacency; time-aware decay with large $\Delta\tau$ reduces its influence.
\end{itemize}

\paragraph{Temporal periodicity (calendar effects).}
\begin{itemize}
\item \textbf{Food delivery.} Saturday-evening orders repeat weekly: index-only can’t align “Saturday 7pm” with earlier Saturdays to anticipate replenishment.
\item \textbf{Music.} A user listens to \emph{jazz} every weekday morning commute and \emph{EDM} on Friday nights. Index-only mixes these regimes; time-aware angles align by time-of-day/day-of-week.
\end{itemize}

\subsubsection*{Why Time-Only RoPE is insufficient}

Similar values of $\Delta\tau$ can mask very different \emph{index} separations, which carry evidence of topical drift.
\begin{itemize}
\item \textbf{Gaming.}
\begin{itemize}
\item  ($\Delta\tau{=}1$ hour, $\Delta i{=}1$): user has a good likelihood on resuming the same \emph{builder} genre;
\item  ($\Delta\tau{=}1$ hour, $\Delta i\gg 1$): multiple \emph{shooter} sessions intervened; next-item is likely to follow shooter cadence.
\end{itemize}

\item \textbf{Video.}
\begin{itemize}
\item  ($\Delta\tau{=}7$ days, $\Delta i{=}1$): likely re-engage the \emph{same creator/topic}.
\item  ($\Delta\tau{=}7$ days, $\Delta i\gg 1$): dozens of intervening clips (cooking, travel); topical center moved.
\end{itemize}

\end{itemize}

\begin{figure*}[t]
  \centering
  \begin{subfigure}{0.45\linewidth}
    \includegraphics[width=\linewidth, trim = {0mm 30mm 30mm 0mm}, clip]{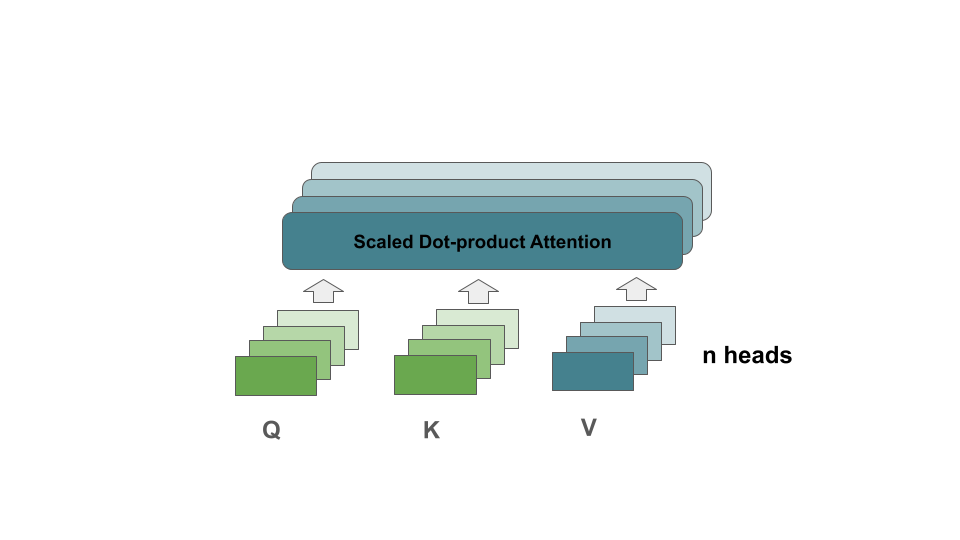}\caption{Index only}\label{fig:index_only}
  \end{subfigure}\hfill
    \begin{subfigure}{0.45\linewidth}
    \includegraphics[width=\linewidth, trim = {0mm 30mm 30mm 0mm}, clip]{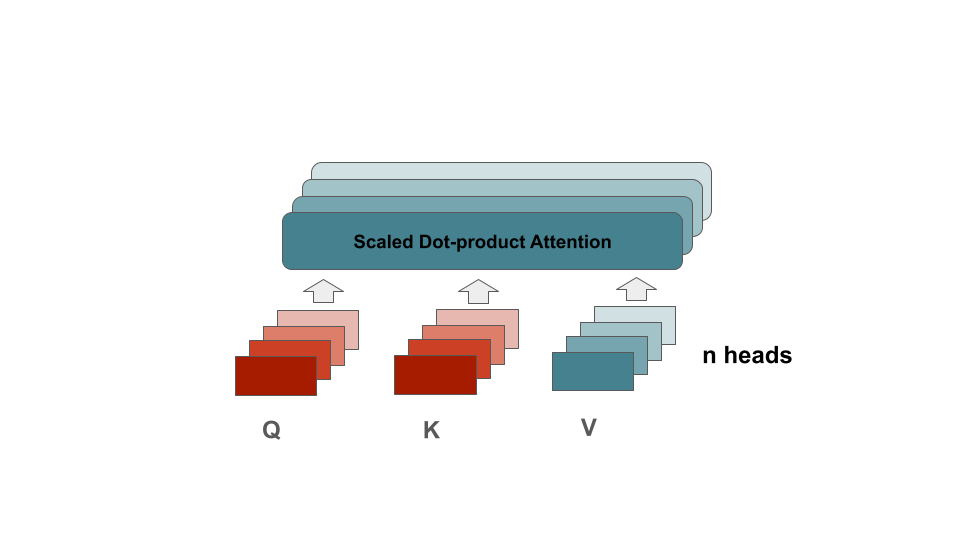}\caption{Time only}\label{fig:time_only}
  \end{subfigure}\hfill
  \begin{subfigure}{0.33\linewidth}
    \includegraphics[width=\linewidth, trim = {50mm 30mm 30mm 50mm}, clip]{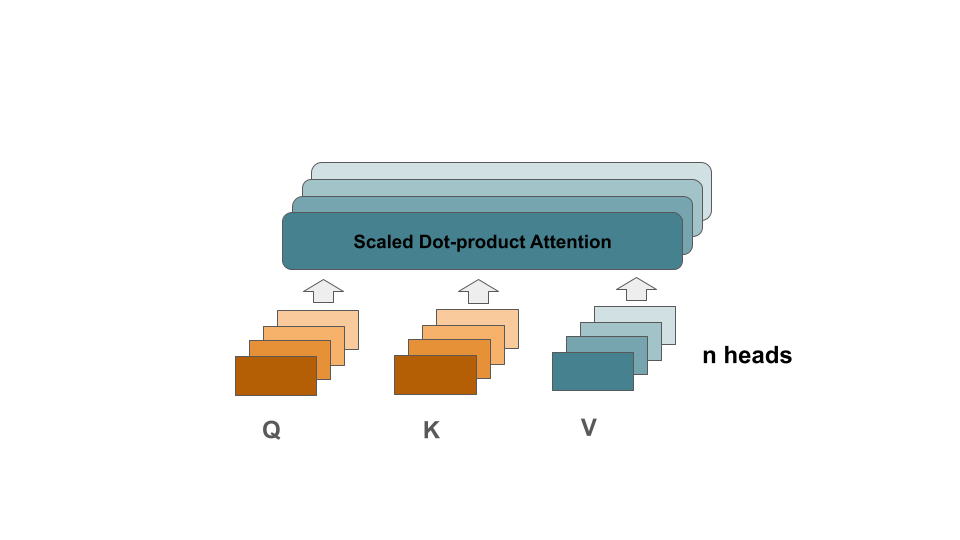}\caption{Early fusion}\label{fig:early_fusion}
       \end{subfigure}\hfill
  \begin{subfigure}{0.33\linewidth}
    \includegraphics[width=\linewidth, trim = {50mm 30mm 30mm 50mm}, clip]{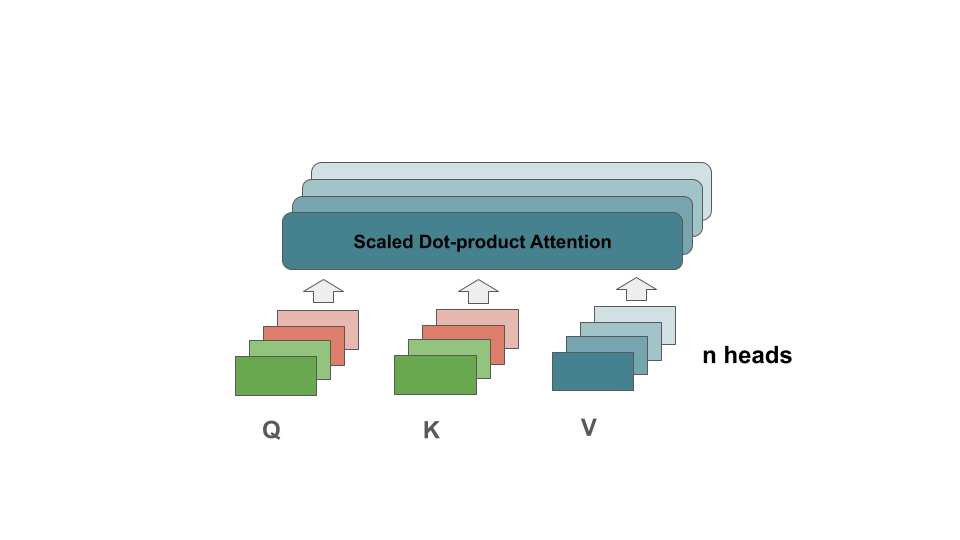}\caption{Split by Head}\label{fig:split_by_head}
  \end{subfigure}\hfill
  \begin{subfigure}{0.33\linewidth}
    \includegraphics[width=\linewidth, trim = {50mm 30mm 30mm 50mm}, clip]{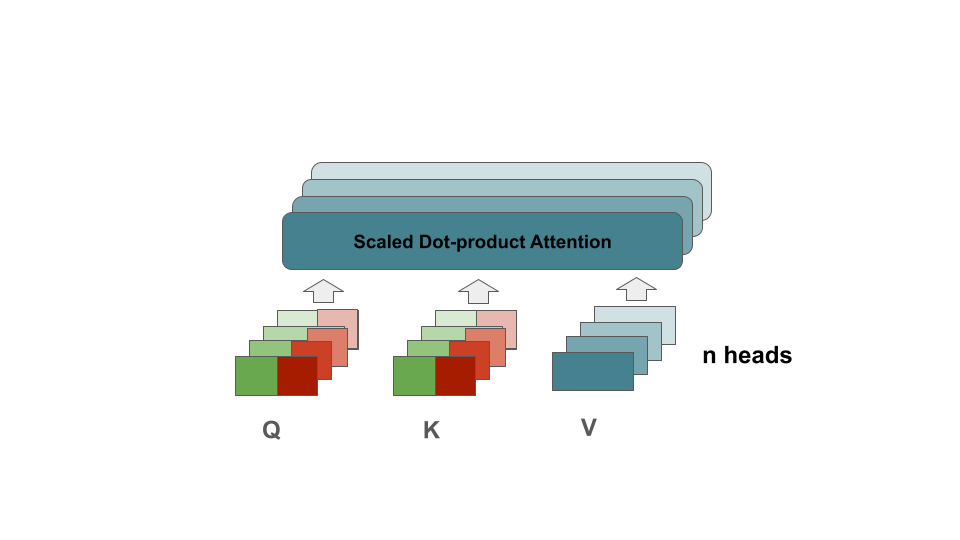}\caption{Split by Dim}\label{fig:split_by_dim}
  \end{subfigure}
  \caption{First row: single source RoPE. Second row: TO-RoPE variants which can incorporate both index and time information}
  \label{fig:torope_variants}
\end{figure*}

%===================================================
\section{Methodology}

In this section, we discuss how to effectively encode both temporal and sequential information in RoPE. We first present a unified formulation for the proposed \emph{Time-and-Order RoPE} (TO-RoPE), which is a family of fusion methods for time and index. Then we consider multiple instantiations of TO-RoPE (early fusion, split-by-dim, split-by-head), and discuss their relative strengths.
\begin{itemize}
\item \textbf{Early Fusion:} a variant couples them \emph{inside} each rotary plane by adding angles
\item \textbf{Split-by-dimension:} dedicate some rotary planes to index ($\theta{=}\theta^{\mathrm{pos}}$) and others to time ($\theta{=}\theta^{\mathrm{time}}$). This removes per-plane interference to become more stable, and offers an explicit capacity knob (e.g., 70\% time / 30\% index).
\item \textbf{Split-by-head:} specialize heads (some index-only, others time-only) so each head retains full plane capacity; interactions emerge across heads via multi-head mixing.
\end{itemize}

\subsection{Time-and-Order RoPE: A Unified View}
\subsubsection{Formulation}
We denote the discrete index by $i$ and a per-token time scalar by $\tau_i$ (e.g., normalized timestamp). Let an attention head have rotary dimension $d$ with $N=\tfrac{d}{2}$ rotary planes indexed by $k\in\{1,\ldots,N\}$. Each plane rotates an even–odd channel pair of $(Q,K)$ by an angle $\theta_{k}(i)$. \emph{TO-RoPE} expresses $\theta_k$ as a linear fusion (in the angle domain) of index and time contributions:
\begin{equation}
\label{eq:to-rope}
\theta_{k}(i)
=
(1-\lambda_{k})\,\alpha^{p}_{k}\, i\,\omega^{p}_{k}
\;+\;
\lambda_{k}\,\alpha^{t}_{k}\, \tau_i\,\omega^{t}_{k},
\qquad
\lambda_{k}\in[0,1],\;\alpha^{p}_{k},\alpha^{t}_{k}>0.
\end{equation}
Here $\omega^{p}_{k}$ and $\omega^{t}_{k}$ are (fixed or learnable) frequency banks for index and time (e.g., geometric ladders); $\alpha$ are optional (fixed or learnable) positive (e.g., softplus) scales, and $\lambda$ is a (fixed or learnable) gate (sigmoid).

For a pair of tokens $(i,j)$, attention depends only on the \emph{relative phase}
\[
\Delta\theta_k \;\coloneqq\; \theta_k(i)-\theta_k(j)
\;=\; \tilde{\omega}^{p}_{k}\,\Delta i \;+\; \tilde{\omega}^{t}_{k}\,\Delta\tau,
\quad
\Delta i \coloneqq i-j,\;\; \Delta\tau \coloneqq \tau_i-\tau_j,
\]
with \emph{effective frequencies}
\[
\tilde{\omega}^{p}_{k} \;=\; (1-\lambda_k)\,\alpha^{p}_{k}\,\omega^{p}_{k},
\qquad
\tilde{\omega}^{t}_{k} \;=\; \lambda_k\,\alpha^{t}_{k}\,\omega^{t}_{k}.
\]
Thus TO-RoPE is a two-feature linear model in the angle space per plane, where $(\tilde{\omega}^{p}_{k},\tilde{\omega}^{t}_{k})$ governs the plane’s sensitivity to order vs.\ time.

Equation~\eqref{eq:to-rope} can be written by learning the plane-wise effective frequencies directly:
\[
\theta_k(i) \;=\; \tilde{\omega}^{p}_k\, i \;+\; \tilde{\omega}^{t}_k\, \tau_i,
\qquad
\tilde{\omega}^{p}_k,\tilde{\omega}^{t}_k \ge 0,
\]
optionally with a monotone ladder over $k$ to preserve multi-scale coverage.

\subsubsection{Special cases (instantiations).}
As shown in Figure \ref{fig:torope_variants}, the unified form subsumes common designs by constraining $\lambda_k$ (or $\omega_k$):
\begin{itemize}
\item \textbf{Index-only/Time-only}: it becomes index-only RoPE when all $\lambda_k=0$ or time-only RoPE when $\lambda_k=1$.
\item \textbf{Early fusion} (global): $\lambda_k\equiv\lambda$, $\alpha^{(\cdot)}_{k}\equiv\alpha^{(\cdot)}$ for all $k$ (or per-head). Angles add inside each plane; most expressive but susceptible to per-plane interference.
\item \textbf{Split-by-dimension}: hard gate per plane, $\lambda_k\in\{0,1\}$. A subset $\mathcal{P}$ uses index-only ($\omega^{t}_k{=}0$), the complement uses time-only ($\omega^{p}_k{=}0$). Removes within-plane interference; exposes an explicit capacity knob.
\item \textbf{Split-by-head}: tie the same hard gate across planes within head $h$, i.e., $\lambda_{h,k}\equiv\lambda_h\in\{0,1\}$. Heads specialize to index or time; interactions emerge via multi-head mixing.
\item \textbf{Other possible instatiations}: split-by-layer which ties gates by depth ($\lambda_\ell$) or split-by-group which enables KV-group ($\lambda_g$) specialization for GQA \cite{Ainslie2023GQATG}.
\end{itemize}

\subsection{Instantiation: Early Fusion}
\subsubsection{Formulation}
Early fusion couples \emph{index} and \emph{time} \emph{inside} each rotary plane by forming a single angle per plane. 
Let a head have $N=\tfrac{d}{2}$ rotary planes indexed by $k$. For token $i$ (index $i$ and time scalar $\tau_i$), the plane-$k$ angle is
\begin{equation}
\label{eq:ef-theta}
\theta_{i,k}
\;=\;
i\,\omega^p_k \;+\; \tau_i\,\omega^t_k,
\end{equation}
with $\omega^p_k,\omega^t_k>0$ (e.g., geometric ladders). Applying the $2\times2$ rotation $R(\theta_{i,k})$ to $(Q,K)$ yields, for token pair $(i,j)$,
\begin{equation}
\label{eq:ef-score}
s_{ij}^{(k)}
\;\propto\;
\cos\!\Big(\underbrace{(i-j)\,\omega^p_k}_{\text{index term }}
+\underbrace{(\tau_i-\tau_j)\,\omega^t_k}_{\text{time term}}\Big).
\end{equation}
Thus, each plane $k$ behaves as a bivariate cosine kernel over $(\Delta i,\Delta\tau)$, and the head score is the sum $\sum_{k=1}^N s_{ij}^{(k)}$.

Using the trigonometric identity $\cos(a{+}b)=\cos a\,\cos b-\sin a\,\sin b$, the (normalized) similarity in plane $k$ depends on the single difference angle
\[
\Delta\theta_{ij,k}
=\Delta i\,\omega^p_k+\Delta\tau\,\omega^t_k,
\quad
\Delta i=i-j,\;\;\Delta\tau=\tau_i-\tau_j,
\]
and can be written as
\[
\cos(\Delta\theta_{ij,k})
= \underbrace{\cos(\Delta i\,\omega^p_k)\cos(\Delta\tau\,\omega^t_k)}_{\text{constructive term}}
\;-\;
\underbrace{\sin(\Delta i\,\omega^p_k)\sin(\Delta\tau\,\omega^t_k)}_{\text{interference term}}.
\]
This single decomposition makes explicit the two interaction pathways within a single rotary plane and explains both position-conditioned periodicity and interference.

If both gaps are small in the plane’s units, $\lvert \Delta i\,\omega^p_k\rvert\ll 1$ and $\lvert \Delta\tau\,\omega^t_k\rvert\ll 1$, then
$\cos(\Delta i\,\omega^p_k)\!\approx\!1$, $\cos(\Delta\tau\,\omega^t_k)\!\approx\!1$, and the sine terms are negligible, so
$\cos(\Delta\theta_{ij,k})\approx 1$.
Hence high-frequency planes (large $\omega$) spike only under simultaneous locality in index and time, sharpening micro-burst attention. With periodic time channels (e.g., time-of-day), the combined phase realizes patterns like “near a 24h multiple \emph{and} near in index,” enabling calendar-aware recency.

\subsubsection{Discussion: Destructive interference.}
When the gaps disagree (e.g., $|\Delta i\,\omega^p_k|\!\ll\!1$ but $|\Delta\tau\,\omega^t_k|$ is moderate), the cosine product is attenuated while the interference (sine) term grows, driving $\cos(\Delta\theta_{ij,k})$ toward zero or negative values: a large time-induced phase can undo index alignment within the same plane (and vice versa).

If we take a gradient view, one can see that per-plane sensitivities are as follows:
\begin{align}
\frac{\partial}{\partial(\Delta i)}\cos(\Delta\theta_{ij,k})=-\,\omega^p_k\sin(\Delta\theta_{ij,k}),
\\
\frac{\partial}{\partial(\Delta\tau)}\cos(\Delta\theta_{ij,k})=-\,\omega^t_k\sin(\Delta\theta_{ij,k}).
\end{align}
When interference pushes $\Delta\theta_{ij,k}$ near $\pm\frac{\pi}{2}$, $\sin(\Delta\theta_{ij,k})$ dominates and flips sign rapidly across pairs, leading to unstable scale learning and oscillatory updates. This helps explain the empirical stability of split variants we will introduce next, which remove the cross sine-cosine interaction by isolating time and index into disjoint planes or heads.

\subsection{Instantiation: Split-by-Dimension (Planes)}
\label{subsec:split-dim}
Let a head have rotary dimension $d$ with $N=\tfrac{d}{2}$ planes $k\in\{1,\dots,N\}$.
Use a \emph{plane gate} $\lambda_k\in\{0,1\}$ that selects the source per plane:
$\lambda_k=0$ for \emph{pos-only}, $\lambda_k=1$ for \emph{time-only}.
The per-plane angle in the unified TO-RoPE form is
\[
\theta_k(i) \;=\; \lambda_k\, (i\,\omega^p_k)
\;+\; (1-\lambda_k)\,(\tau_i\,\omega^t_k),
\qquad
\omega^{(\cdot)}_k>0,
\]
so that each plane encodes either index or time. Let $P{=}\{k:m_k{=}1\}$ and $T{=}\{k:\lambda_k{=}0\}$ with $|P|{+}|T|{=}N$.

For a query at $i$ and a key at $j$, the rotary similarity in head $h$ decomposes additively across planes:
\begin{equation}
\label{eq:splitdim-score}
s^{(h)}_{ij}
\;=\;
\sum_{k\in P} \big\langle R(\Delta i\,\omega^p_k) q^{(h)}_k,\; k^{(h)}_k \big\rangle
\;+\;
\sum_{k\in T} \big\langle R(\Delta \tau_{ij}\,\omega^t_k) q^{(h)}_k,\; k^{(h)}_k \big\rangle,
\end{equation}
where $\Delta i{=}i{-}j$, $\Delta\tau_{ij}{=}\tau_i{-}\tau_j$, $R(\cdot)$ denotes a $2{\times}2$ rotation on the even-odd channel pair, and $q^{(h)}_k,k^{(h)}_k\in\mathbb{R}^2$ are the plane-$k$ slices of $Q,K$.
Under a random-feature view with energy-normalized planes, \eqref{eq:splitdim-score} behaves like a bank of cosine kernels:
\[
\mathbb{E}[s^{(h)}_{ij}] \;\propto\;
\sum_{k\in P}\cos(\Delta i\,\omega^p_k)
\;+\;
\sum_{k\in T}\cos(\Delta \tau_{ij}\,\omega^t_k).
\]
Capacity allocation on time/index is explicit: one can choose a large fraction $\rho{=}|T|/N$ (e.g., $0.7$) to skew capacity toward time if time is more important for a particular dataset.

If we compare the split-by-dim variant to the early fusion approach, split-by-dim has the following advantages:
\begin{itemize}
\item \textbf{No per-plane interference}: a plane never mixes time and index, preventing destructive phase cancellation.
\item \textbf{Interpretable}: $P$ captures ordinal locality; $T$ captures wall-clock effects; frequency routing is transparent.
\item \textbf{Periodic channels}: dedicate selected planes in $T$ to different periodicity bands (e.g., daily/weekly).
\end{itemize}

\subsection{Instantiation: Split-by-Head}
\label{subsec:split-head}
Partition heads into $H_p$ (pos-only) and $H_t$ (time-only) with 
\[H_p{\cup}H_t{=}\{1,\dots,H\}, H_p{\cap}H_t{=}\emptyset.\]

For $h\in H_p$, every plane uses $\theta^{(h)}_k(i)=i\,\omega^p_k$ and for $h\in H_t$, $\theta^{(h)}_k(i)=\tau_i\,\omega^t_k$. Denote the head-wise attention scores
\[
s^{(h)}_{ij} \;=\; \sum_{k=1}^{N} \big\langle R(\theta^{(h)}_k(i)-\theta^{(h)}_k(j))\, q^{(h)}_k,\; k^{(h)}_k \big\rangle,
\quad
h\in H_p\cup H_t,
\]
and the multi-head aggregation $S_{ij} = \frac{1}{\sqrt{d}}\sum_{h=1}^{H} s^{(h)}_{ij}$ with the usual softmax over $j$.
Modality interactions occur when head outputs are mixed by the output projection $W_o$ and subsequent layers:
\[
\mathrm{out}_i \;=\; \bigg[\;\big\|_{h=1}^{H} \sum_j \mathrm{softmax}_j(S_{ij})\, V^{(h)}_j\bigg] W_o,
\]
so that information from $H_p$ and $H_t$ combines in the concatenation and $W_o$ mixing.

It is straightforward to implement and compatible with SDPA/Flash-Attention \cite{Dao2022FlashAttentionFA, Dao2023FlashAttention2FA}. In this variant, each head keeps all $N$ planes for its modality (i.e., index or time) , enabling fine frequency coverage within the head. Some heads could focus on short-range order (pos), while others focus on long-range recency/periodicity (time). 

Consdier the streaming example for a user with watchpath: \emph{S1E1 $\rightarrow$ S1E2 $\rightarrow$ S1E3} (tight), then two weeks later movie night.  
Index heads emphasize continuation (E3 $\rightarrow$ E4) despite the later movie, while time heads temper influence from long-paused items and emphasize recent modality switches. The (aggregated) multi-head attention score would try to strike a balance between these tensions in a data-driven manner. By constrast, index-only tends to over-pull the paused show after long gaps and time-only might ignore episodic order.

\begin{table*}[t]
\centering
\small
\renewcommand{\arraystretch}{1.15}
\begin{tabularx}{\textwidth}{@{}lXXX@{}}
\toprule
\textbf{Aspect} &
\textbf{Early fusion} &
\textbf{Split-by-dim (planes)} &
\textbf{Split-by-head} \\
\midrule

\textbf{Geometry effect on $Q/K$} &
Modifies $Q/K$ \emph{jointly} (phase interactions within a plane). &
Changes $Q/K$ but \emph{separately} per plane (no interference). &
Changes $Q/K$ but \emph{separately} per head (no interference). \\
\addlinespace
\textbf{Short-term bursts} &
Good: high-freq planes peak only when $\Delta i$ \emph{and} $\Delta\tau$ are small. &
Good, via separate planes (less within-plane sharpness). &
Good, via specialized heads (needs cross-head mixing). \\
\addlinespace
\textbf{Long-range rhythms} &
Can be suppressed if poorly scaled. &
Good (allocate low-freq planes to time). &
Good (time heads capture long gaps). \\
\addlinespace
\textbf{Periodicity (TOD/DOW/season)} &
Pos-conditioned periodicity can emerge within planes. &
Good: dedicate planes to periodic bands. &
Good: dedicate heads to periodic bands. \\
\addlinespace
\textbf{Robustness} &
Weaker due to the angle interference. &
Robust; explicit capacity split limits interference. &
Robust; head specialization contains errors. \\
\bottomrule
\end{tabularx}
\caption{Comparison of three RoPE-based strategies to combine discrete index and wall-clock time in generative recommendation. Early fusion modifies the $Q/K$ geometry jointly, enabling within-plane interactions but requiring careful scale calibration. Split-by-dim allocates rotary \emph{planes}; split-by-head allocates entire \emph{heads}.}
\label{tab:rope-methods}
\end{table*}

%============================================                                     

\section{Experiments}

\label{sec:experiments}

In this section, we empirically evaluate \emph{TO-RoPE} across public and proprietary datasets and organize our study around three research questions:
\begin{itemize}
\item{RQ1 (Comparative performance).} How does time–and–order RoPE compare to popular alternatives such as \emph{relative-bias} positional encodings (e.g., HSTU-style relative bias \cite{Zhai2024ActionsSL}) and \emph{absolute} schemes (sinusoidal, learned tables) in terms of prediction accuracy?
\item{RQ2 (Qualitative Analysis \& Case studies).} What factors might explain the effectiveness of TO-RoPE compared to alternatives (e.g., absolute time or relative bias)

\item{RQ3 (Capacity allocation).} How does capacity allocation/split ratio between time and index (by \emph{dimension} and by \emph{head}) affect model performance across different datasets?

\end{itemize}

\begin{table*}[htbp]
\centering
\caption{Performance comparison between different PE variants on our proprietary dataset. Boldface denotes the highest value while underline indicates the second best result.}
\label{tab:rope_variants}
\renewcommand{\arraystretch}{1.2} % Increases the vertical spacing between rows
\begin{tabular}{l|ccccc}
\hline
\textbf{Model} & \textbf{HR@10} & \textbf{NDCG@10} & \textbf{HR@50} & \textbf{NDCG@50} & \textbf{HR@300} \\
\hline

% --- Baseline Section ---
\textit{Baseline} & & & & & \\
Index APE & 0.5493 & 0.3805 & 0.7588 & 0.4273 & 0.9245 \\
Index APE + time APE & 0.5510 & 0.3818 & 0.7605 & 0.4286 & 0.9256 \\
Relative Bias(HSTU-style) & 0.5513 & 0.3820 & 0.7604 & 0.4287 & 0.9255 \\
\hline

\textit{RoPE} & & & & & \\
Index only RoPE & 0.5537 & 0.3841 & 0.7612 & 0.4305 & 0.9253 \\
Time only RoPE & 0.5568 & 0.3865 & 0.7653 & 0.4331 & 0.9278 \\
Index only RoPE + time APE & 0.5547 & 0.3848 & 0.7627 & 0.4313 & 0.9262 \\
\hline

% --- Early Fusion Section ---
\textit{Time-and-Order RoPE} & & & & & \\
Early fusion & 0.5562 & 0.3855 & 0.7648 & 0.4321 & 0.9277 \\
Split head & \textbf{0.5582} & \textbf{0.3875} & \textbf{0.7663} & \textbf{0.4340} & \textbf{0.9281} \\
Split dim & \underline{0.5582} & \underline{0.3874} & \underline{0.7663} & \underline{0.4339} & \underline{0.9281} \\
\hline
\end{tabular}
\end{table*}

\subsection{Experiment Setup}
\label{ssec:setup}

\subsubsection{Datasets.}
\label{par:datasets}
We evaluate using two distinct datasets:
\begin{itemize}
\item  \textbf{MovieLens-20M:} This is the MovieLens dataset consisting of user, movie and corresponding ratings and timestamps. Following established preprocessing protocols, we apply 5-core filtering, retaining only users and movies with at least five interactions.
\item \textbf{Proprietary Dataset:} A large-scale, in-house dataset from an online gaming platform, which consists of \emph{tens of millions} of users and their interations with content on this platform over one year. This dataset presents more challenges from real-world scenario, including large number of users, longer sequence lengths and a much larger action space.
\end{itemize}

\subsubsection{Models and Baselines.}
\label{par:models}
For both public and proprietary datasets, we experiment with different positional encoding methods using the same generative recommendation model backbone, which is an autoregressive decoder-only (GPT-2 style \cite{Radford2019LanguageMA}) with $12$ transformer layer and about 85M parameters. We consider the following baselines and TO-RoPE variants:
\begin{itemize}
\item Absolute Positional Encoding (APE): learnable position embedding based on index (GPT-2 style \cite{Radford2019LanguageMA}) or bucktized time.
\item Relative Positional Encoding (RPE): similar to T5-bias \cite{Raffel2019ExploringTL} and HSTU \cite{Zhai2024ActionsSL}, where we add time-interval and index based relative bias to attention logits.
\item Single-Source Rotary Positional Encoding: index-only RoPE and time-only RoPE.
\item TO-RoPE instantiations: early fusion, split-by-head and split-by-dim.
\end{itemize}

\subsubsection{Evaluation Metrics.}
\label{par:metrics}
For evaluation on MovieLens-20M, we follow the convention of leave-one-out strategy \cite{Kang2018SasRec} to split train, validation and test sets. For our proprietary dataset, we leave out the last day for testing, the second last day for validation, and the rest for training.
For both datasets, we use top-k Hit Rate ($HR@k$) and Normalized Discounted Cumulative Gain ($NDCG@k$) as metrics to evaluate different methods, where $k \in \{10, 50, 200\}$ for ML-20M dataset and $k \in \{10, 50, 300\}$ for our proprietary dataset.

\subsubsection{Implementation Details.}
\label{par:hyperparams}
For MovieLens-20M, we set the maximum input sequence length to $50$. We adopt the commonly used sequential recommendation practices \cite{Kang2018SasRec} in the literacture including full shuffle and multi-epoch training. We use a learning rate of $0.001$, batch size of 128 and dropout ratio of $0.2$. For proprietary dataset, the maximum sequence length is set to $1024$. 

% We employ a stream training setting, where max tokens per batch is set to 32k. We use a learning rate of 0.0001 and a linear warm-up of 5000 steps.

\subsection{RQ1 — Comparative Performance}
\textbf{Question.} How does time–and–order RoPE compare to \emph{relative-bias} positional encodings (e.g., HSTU-style rab \cite{Zhai2024ActionsSL}) and \emph{absolute} schemes (sinusoidal, learned tables) on top-$K$ metrics?

\noindent\textbf{Result.} We compare different positional encoding setup on both public and proprietary datasets. For more comprehensive comparison, for index-only baselines, we also include different variants that have both order and time information present. Results are shown in Table~\ref{tab:rope_variants} for proprietary data and Table~\ref{tab:pe_variants_ml20m} for Movielens-20M data. First, RoPE variants consistantly outperform their APE (Absolute Position Encoding) and RPE (Relative Position Encoding) counterparts. Second, time only RoPE performs better than index only RoPE on both datasets, incidating the usefulness of time information under sequential recommendation settings. Third, early fusion method sometimes fails to outperform time-only RoPE (e.g., on our proprietary dataset). This indicates the interference effect in RoPE's rotation with fused time and index angle is not always constructive, which validates our motivation to explore and develop better instantiations under TO-RoPE framework. Fourth, by careful design of the way to combine time and index in RoPE, our proposed split-by-dim and split-by-head approaches outperform all other positional encoding variants on both datasets.

\begin{table*}[htbp]
\centering
\caption{Performance comparison between different PE variants on the ML-20M dataset. Boldface denotes the highest value while underline indicates the second best result.}
\label{tab:pe_variants_ml20m}
\renewcommand{\arraystretch}{1.2} % Increases the vertical spacing between rows
\begin{tabular}{l|cccccc}
\hline
\textbf{Model} & \textbf{HR@10} & \textbf{HR@50} & \textbf{HR@200} & \textbf{NDCG@10} & \textbf{NDCG@50} & \textbf{NDCG@200} \\
\hline

% --- Baseline Section ---
\textit{Baseline} & & & & & & \\
Index APE & 0.3318 & 0.5729 & 0.7669 & 0.2007 & 0.2540 & 0.2835 \\
Index APE + time APE & 0.3335 & 0.5739 & 0.7682 & 0.2023 & 0.2554 & 0.2849 \\
Relative Bias(HSTU-style) & 0.3341 & 0.5727 & 0.7668 & 0.2023 & 0.2550 & 0.2845 \\
\hline
\textit{RoPE} & & & & & & \\
Index only RoPE & 0.3347 & 0.5737 & 0.7681 & 0.2026 & 0.2555 & 0.2850 \\
Time only RoPE & 0.3341 & 0.5746 & 0.7697 & 0.2027 & 0.2559 & 0.2856 \\
Index only RoPE + time APE & 0.3353 & 0.5746 & 0.7690 & 0.2031 & 0.2561 & 0.2856 \\
\hline

% --- Early Fusion Section ---
\textit{Time-and-Order RoPE} & & & & & & \\
Early fusion & 0.3362 & 0.5765 & \textbf{0.7790} & 0.2037 & 0.2569 & 0.2864 \\
Split head & \underline{0.3388} & \underline{0.5767} & 0.7709 & \underline{0.2048} & \underline{0.2574} & \underline{0.2869} \\
Split dim & \textbf{0.3406} & \textbf{0.5791} & \underline{0.7711} & \textbf{0.2059} & \textbf{0.2587} & \textbf{0.2879} \\
\hline
\end{tabular}
\end{table*}

\subsection{RQ2 - Understanding the effect the RoPE on Modeling User sequence}
\textbf{Question.} Why TO-RoPE tend to be more effective compared to alternatives (e.g., absolute time or relative bias)?

\begin{figure}[h]
 \centering
  \includegraphics[width=0.86\columnwidth]{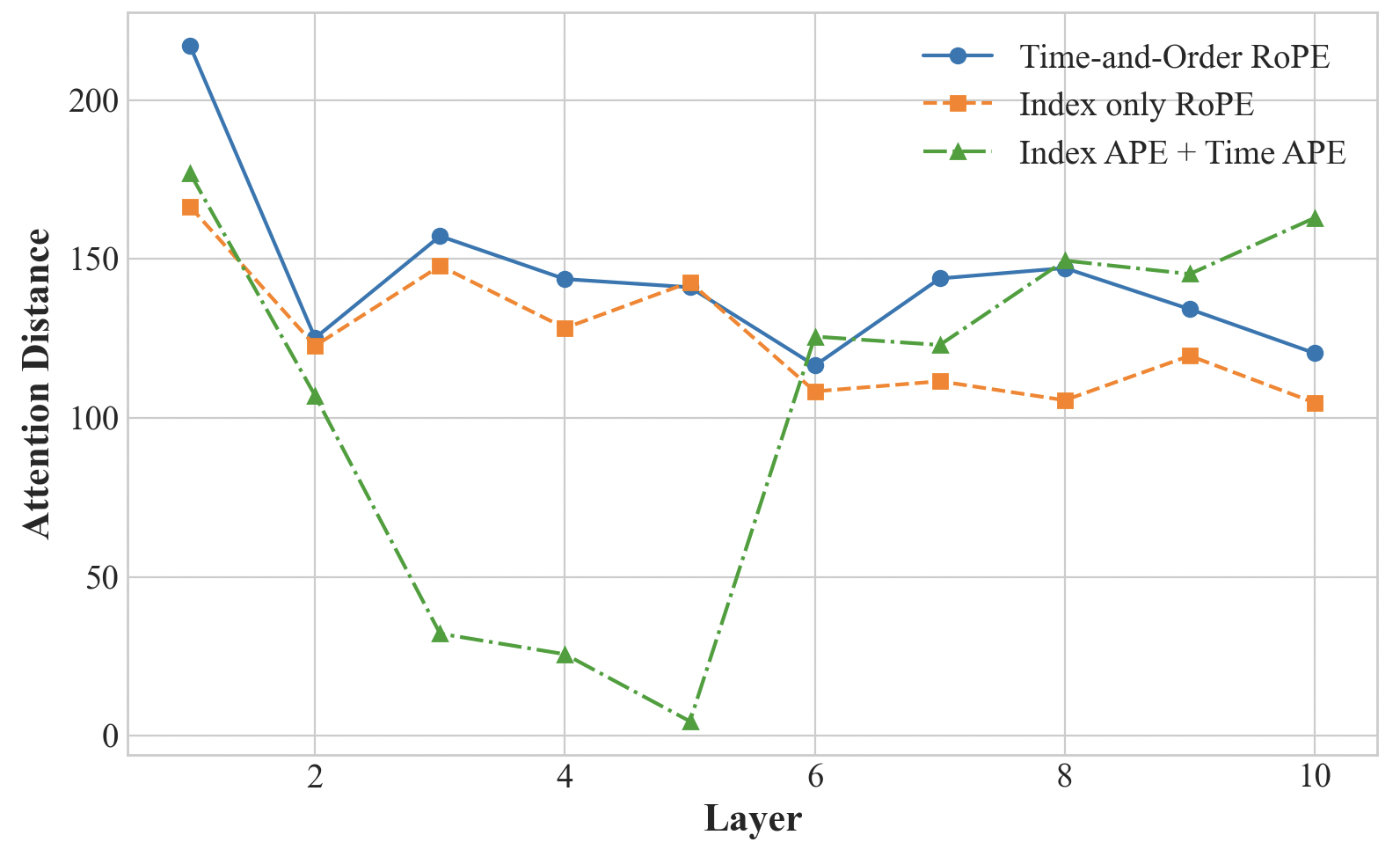}
 \caption{Attention distances for different positonal encoding variants. We measure the average attention distance by weighting query-key distance with attention probabilities.}
 \label{fig:attention_distance}
 \end{figure}

\begin{figure}[h]
 \centering
  \includegraphics[width=0.86\columnwidth]{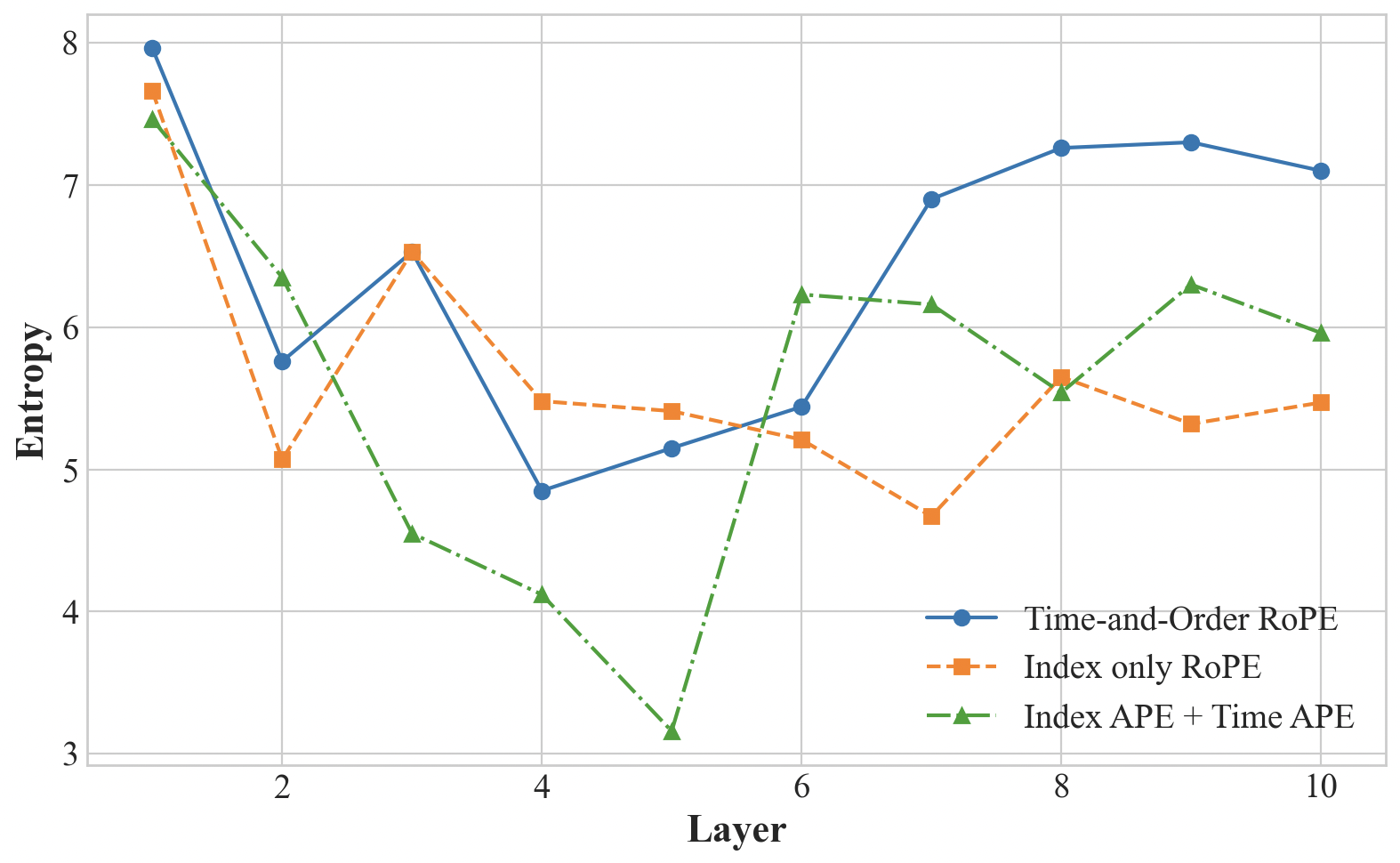}
 \caption{Attention entropy for different positonal encoding variants. Higher entropy indicates more tokens are involved in attention.}
  \label{fig:attention_entropy}
 \end{figure}

\noindent\textbf{Result.} We evaluate on the real user action sequence in our proprietary dataset to help understand the mechanism behind TO-RoPE. We look into the attention pattern for different position encoding variants, and measure the attention distance and entropy on $1k$ randomly sampled users who have at least $500$ interactions. We average the result across the user and head dimention and report the average metrics per layer. Attention distance works as a good indicator for range of attention interaction and entropy is good for measuring the number of tokens involved. As can be seen from Figure~\ref{fig:attention_distance} and Figure~\ref{fig:attention_entropy}, compared with APE, RoPE variants show significant increase in both attention distance and entropy in middle layers. Compared with index only RoPE, our TO-RoPE approach tend to exhibit higher distance and entropy across all layers. These results indicate TO-RoPE is more capable of attending to \emph{longer-range} and \emph{more} interactions/items in the user history. We speculate such improved context retrieval ability contributes to the performance improvement.

\subsection{RQ3 - Capacity Allocation}
\textbf{Question.} How does capacity allocation/split ratio between time and index (by \emph{dimension} and by \emph{head}) affect model performance across different datasets?

\noindent\textbf{Result.} As shown in the previous results, the split-by-dim and split-by-head variants tend to be stronger variants which outperform early fusion variant. In this subsection, we invetigate how sensitive these variants are w.r.t different ratios of capacity (head/dim) allocation. Concretely, we vary the split ratios in $\{0.1, 0.3, 0.5, 0.9\}$ for these two methods on our propritary dataset as well as ML-20M, and report the comparison in Table \ref{tab:rope_variants_adjusted}. We observe that split-by-head and split-by-dim with ratio $0.3$ - $0.5$ could typically provide decent performance, and such ratio might be a safe default choice. Practically, one can tune the split-ratio as a hyperparameter for their specific data and use cases, to achieve capacity (head/dim) allocation that maximizes information utilization.

\begin{table*}[htbp]
\centering
\caption{Comparison between different frequency allocation strategies on different datasets. Boldface denotes the highest value while underline indicates the second best result. Higher split ratio indicates higher time side contribution.}
\renewcommand{\arraystretch}{1.2}
\label{tab:rope_variants_adjusted}
\begin{tabular}{l|cccc|cccc}
\hline
\multirow{2}{*}{\textbf{Model}} & \multicolumn{4}{c|}{\textbf{Proprietary}} & \multicolumn{4}{c}{\textbf{ML-20M}} \\
\cline{2-9}
& \textbf{HR@10} & \textbf{NDCG@10} & \textbf{HR@50} & \textbf{NDCG@50} & \textbf{HR@10} & \textbf{NDCG@10} & \textbf{HR@50} & \textbf{NDCG@50} \\
\hline

% --- Split By Head Section ---
\textit{Split By Head} & & & & & & & & \\
Split head 0.1 & 0.5580 & 0.3873 & 0.7662 & 0.4338 & 0.3367 & 0.2042 & 0.5772 & 0.2574 \\
Split head 0.3 & \textbf{0.5582} & \textbf{0.3875} & \underline{0.7663} & \textbf{0.4340} & \underline{0.3388} & \underline{0.2048} & 0.5767 & \underline{0.2574} \\
Split head 0.5 & 0.5579 & 0.3871 & 0.7662 & 0.4336 & 0.3370 & 0.2039 & 0.5760 & 0.2567 \\
Split head 0.9 & 0.5577 & 0.3871 & 0.7662 & 0.4336 & 0.3370 & 0.2038 & 0.5765 & 0.2568 \\
\hline

% --- Split By Dim Section ---
\textit{Split By Dim} & & & & & & & & \\
Split dim 0.1 & 0.5572 & 0.3866 & 0.7654 & 0.4331 & 0.3338 & 0.2026 & 0.5734 & 0.2556 \\
Split dim 0.3 & 0.5578 & 0.3871 & 0.7660 & 0.4336 & 0.3350 & 0.2032 & 0.5742 & 0.2560 \\
Split dim 0.5 & \underline{0.5582} & \underline{0.3874} & \textbf{0.7663} & \underline{0.4339} & \textbf{0.3406} & \textbf{0.2059} & \textbf{0.5791} & \textbf{0.2587} \\
Split dim 0.9 & 0.5574 & 0.3868 & 0.7661 & 0.4334 & 0.3371 & 0.2037 & \underline{0.5774} & 0.2569 \\
\hline
\end{tabular}
\end{table*}

\section{Conclusion}

Rotary Position Embedding (RoPE) has become a default choice in modern LLMs because it can inject relative, multi-scale positional structure into attention, while remaining kernel-friendly and cache-friendly. This paper argues that these desirable properties also make RoPE particularly well-suited for modeling user histories in transformer-based generative recommendation. In recommendation scenarios, as temporal information is an important dimension in additional to positional/sequential information, we pose the central question: how to enable RoPE to encode both time and index effectively. To answer this question, we present a unified Time-and-Order RoPE (TO-RoPE) framework that fuses sources in the angle/frequency domain. With this unified view, we discuss expressivity–stability trade-offs and how this framework provides principled knobs for frequency capacity allocation. To instantiate the framework, three practical designs are derived and presented: (i) early fusion (angles add within a plane), (ii) split-by-dim (dedicated planes), and (iii) split-by-head (head specialization).  On both public and proprietary/industrial datasets, we find properly designed TO-RoPE (especially split-by-dim or split-by-head) consistently outperforms absolute embeddings and relative-bias baselines. Our findings demonstrate the potential of TO-RoPE as an effective building block for modeling user behavior sequence in recommendation systems.

\clearpage 

\bibliographystyle{ACM-Reference-Format}
\bibliography{reference}

\end{document}